\documentclass[twocolumn,floatfix,pra,showpacs,aps,superscriptaddress]{revtex4-1}
   
\usepackage{amsmath}
\usepackage{amsbsy}
\usepackage{amsthm}
\usepackage{amssymb}
\usepackage{latexsym}
\usepackage[dvips]{color,graphicx}

\begin{document}

\title{Superfluid to normal phase transition in strongly correlated bosons in \\ two
and three dimensions}

\author{Juan Carrasquilla}
\author{Marcos Rigol}
\affiliation{Department of Physics, Georgetown University, Washington, D.C. 20057, USA}

\affiliation{Physics Department, The Pennsylvania State University, 104 Davey Laboratory, University Park, Pennsylvania 16802, USA}

\begin{abstract}
Using quantum Monte Carlo simulations, we investigate the finite-temperature phase
diagram of hard-core bosons ($XY$ model) in two- and three-dimensional lattices. To 
determine the phase boundaries, we perform a finite-size-scaling analysis of the 
condensate fraction and/or the superfluid stiffness. We then discuss how these phase 
diagrams can be measured in experiments with trapped ultracold gases, where the systems 
are inhomogeneous. For that, we introduce a method based on the measurement of the 
zero-momentum occupation, which is adequate for experiments dealing with both 
homogeneous and trapped systems, and compare it with previously proposed approaches.
\end{abstract}
\pacs{64.60.-i, 67.85.-d, 03.75.Hh, 02.70.Ss}
\maketitle

\section{Introduction}\label{intro}

The description of strongly correlated bosonic systems is of fundamental interest in largely
diverse physical situations ranging from low-temperature experiments with superfluid helium 
\cite{daunt_smith_54} to Josephson-junction arrays \cite{bruder_fazio_schon_93}, as well as
magnetic insulators \cite{giamarchi_ruegg_tchernyshyov_08} and ultracold gases in optical 
lattices \cite{bloch_dalibard_review_08,cazalilla_review_11}. The latter systems offer an 
unparalleled playground to study fundamental models widely considered in statistical and 
condensed-matter physics. This is because of the high degree of control over the experimental 
parameters that determine the Hamiltonian describing the system. In particular, the Bose-Hubbard 
model \cite{fisher_weichman_89,jaksch_bruder_98} has been experimentally realized in 
one \cite{stoferle_moritz_04}, two \cite{spielman_phillips_07,jimenez_spielman_10}, and three 
dimensions \cite{greiner_mandel_02a}, where the superfluid to Mott-insulator transition has been 
observed. Even though it has received less attention, the superfluid to normal transition in 
the Bose-Hubbard model has been investigated experimentally in three dimensions \cite{trotzky_troyer_10}, 
while in two dimensions it has been realized in the form of a two-dimensional lattice of 
Josephson-coupled Bose-Einstein condensates \cite{schweikhard_cornell_07,trombettoni_sodano_05}, 
as well as in experiments with ultracold atoms in optical lattices \cite{zhang_chin_12}.
 
Although experiments with ultracold atoms on optical lattices are in some respects almost 
ideal realizations of model Hamiltonians of interest, significant complications arise because
of the presence of a confining potential, which leads to the coexistence of different phases in 
a single experimental setup \cite{batrouni_rousseau_02,rigol_09}. Furthermore, the mesoscopic 
size of the system in combination with the inhomogeneity induced by the trapping potential produces 
a rounding off of the otherwise sharp features present in an infinite homogeneous system in the 
critical region \cite{rigol_03,wessel_04,campostrini_vicari_10a,campostrini_vicari_10b}. Thus 
the understanding and assessment of criticality in such systems remains a challenging task.
 
The emergence of sharp features in the momentum distribution as obtained from time-of-flight 
images has been frequently associated to the emergence of superfluidity \cite{greiner_mandel_02a,chin_ketterle_06,xu_ketterle_06,tung_cornell_10,clade_phillips_09}. 
However, this association may not be accurate because sharp peaks in the momentum 
distribution already appear in the normal state, due to an increasing correlation length when 
approaching a critical regime \cite{kashurnikov_svistunov_02,diener_ho_07,kato_trivedi_08}. 
More recently, new schemes to detect criticality in trapped systems have been proposed. 
In some of those studies, a detailed analysis of the momentum distribution was used to define 
criteria that allow one to extract reliable estimations of the critical points from 
time-of-flight images \cite{diener_ho_07,trotzky_troyer_10,pollet_svistunov_10}. In addition 
to time-of-flight images, high-resolution \textit{in situ} imaging of the density profile of 
trapped systems has become a powerful instrument with which one can also study phase diagrams 
of strongly correlated systems and quantum criticality. Numerous theoretical and experimental 
studies based on this idea have been carried out for systems in the presence of an optical lattice 
\cite{folling_widera_06,zhou_trivedi_09,gemelke_Zhang_09,ho_zhou_10,nascimbene_salomon_10,
ma_troyer_10,bakr_peng_10,sherson_weitenberg_10, zhang_chin_11,fang_wang_11,hazzard_mueller_11} 
and in the absence of it \cite{kruger_dalibard_07,nascimbene_salomon_nature_10,hung_chin_11,duchon_trivedi_11}.

One important aspect that determines the nature of the quantum phases and their associated order 
parameters is the dimensionality $d$. Mermin {\it et al.} rigorously proved that at any 
nonzero temperature, continuous symmetries cannot be spontaneously broken in systems with 
sufficiently short-range interactions in dimensions $d\le2$ \cite{mermin_wagner_66,hohenberg_67}. 
This implies that, at finite temperature, Bose-Einstein condensation (BEC) cannot occur in one and two 
dimensions. Two-dimensional Bose systems, however, are marginal in the sense that fluctuations 
are strong enough to destroy the fully ordered state but are not so strong as to suppress 
superfluidity. Thus critical behavior develops in the Berezinskii-Kosterlitz-Thouless (BKT) 
transition \cite{berezinskii_72,kosterlitz_thouless_73}, where a superfluid phase with 
quasi-long-range order competes with thermal fluctuations and induces a continuous phase 
transition to the normal fluid as the temperature is increased. In addition to low-temperature 
superfluidity, long-range order can develop at zero temperature in two dimensions. On the 
other hand, in three dimensions, the superfluid transition is accompanied by the appearance of 
true long-range order, implying that the system also exhibits Bose-Einstein condensation. Such 
a transition, which belongs to the three-dimensional $XY$ universality class, is well understood 
in the sense that the critical exponents have been determined experimentally and theoretically 
with remarkably high accuracy in many different physical contexts 
\cite{li_teitel_89,salamon_howson_93,overend_lawrie_94,campostrini_vicari_01,
campostrini_vicari_06,burovski_svistunov_06,hasenbusch_torok_99}.
 
Here, we focus our study on the superfluid to normal transition in a system of strongly 
interacting bosons in two- and three-dimensional lattices. Specifically, we consider 
the Bose-Hubbard model in the limit of infinite on-site repulsion, i.e., the hard-core boson limit. 
We use exact quantum Monte Carlo simulations to compute the finite-temperature phase diagram as a
function of chemical potential. Accurate results are obtained through finite-size scaling of the 
condensate fraction and/or the superfluid stiffness obtained from our simulations. We also determine
the mean-field phase diagram, which is qualitatively correct but quantitatively quite different from the 
exact results. We then proceed to study the superfluid to normal phase transition in two and three 
dimensions in the presence of a confining potential, which is required to describe experiments with 
ultracold gases. We introduce a method to determine the critical temperature, for any given density, 
that is based on the measurement of the zero-momentum occupation as a function of temperature. This 
method is in principle adequate for experiments dealing with both homogeneous and trapped systems. 
Furthermore, we compare our approach to other recently proposed schemes based on the {\it in situ} 
density images \cite{zhou_trivedi_09} as well as on the shape of the low-momentum part of the 
momentum distribution \cite{pollet_svistunov_10}.

The paper is organized as follows. In Sec.~\ref{model}, we introduce the model and its phase diagram 
in two and three dimensions supplemented with the mean-field calculations. Section~\ref{homog} is 
devoted to the discussion of the techniques to obtain the phase boundaries. In Sec.~\ref{trapped}, 
we discuss the possibility to have Bose-Einstein condensation in trapped two-dimensional systems 
as well as the methods to determine the phase boundaries from experimentally accessible quantities. 
Finally, in Sec.~\ref{conclusions}, we draw our conclusions. 

\section{Model and phase diagram} \label{model}

We consider a system of hard-core bosons on a $d$-dimensional lattice with $L^d$ sites. The Hamiltonian 
can be written as
\begin{eqnarray}
\hat{H} &=& -t\sum_{\langle i,j \rangle} \left(\hat{a}^\dagger_{i}\hat{a}^{}_{j} + \textrm{H.c.}\right)  
-\sum_{i}\mu_i \hat{n}_{i} \,\,, \label{hcb}
\end{eqnarray}
where $\hat{a}^\dagger_{i}$ ($\hat{a}_{i}$) is the boson creation (annihilation) 
operator at a given site $i$, and $\hat{n}_{i} = \hat{a}^\dagger_{i} \hat{a}^{}_{i}$ 
is the local particle number operator. The hard-core boson creation and 
annihilation operators satisfy the constraint $\hat{a}^{\dagger 2}_i=\hat{a}^{2}_{i}=0$, which 
forbids multiple occupancy of lattice sites. The first term in Eq.~\eqref{hcb} is the kinetic 
energy, where $t$ is the hopping amplitude between neighboring sites $i$ and $j$ 
($\langle i,j \rangle$). In experiments involving ultracold gases, a trap is required to 
confine the atoms. The effect is taken into account in the second term that
contains $\mu_i=\mu-V_0r_i^{2}$, where $V_0$ is its strength and $\mu$ is the overall 
chemical potential. $r_i$ is the distance from site $i$ to the center of the trap. In what 
follows, positions will be given in units of the lattice spacing $a$ and the energy will be given 
in units of the hopping amplitude $t$. 

We recall that the Hamiltonian in Eq.~\eqref{hcb} can be exactly mapped to the extensively 
studied quantum $XY$ model 
\cite{matsubara_matsuda_56}
\begin{eqnarray}
\hat{H} &=& -2t\sum_{\langle i,j \rangle} \left( S^{x}_{i} S^{x}_{j}+ S^{y}_{i} S^{y}_{j}\right)  
-\sum_{i}\mu_i S^{z}_{i} \,\,, \label{xym}
\end{eqnarray}
where $S^{\alpha}_i$ is the $\alpha$th component of the spin-$1/2$ spin operator at site 
$i$. In the spin language, the term proportional to $t$ describes a ferromagnetic 
exchange interaction, while the one proportional to $\mu_i$ describes a magnetic field 
in the $z$-direction at site $i$. 

\begin{figure}[!t]
\includegraphics[width=0.35\textwidth,angle=-90]{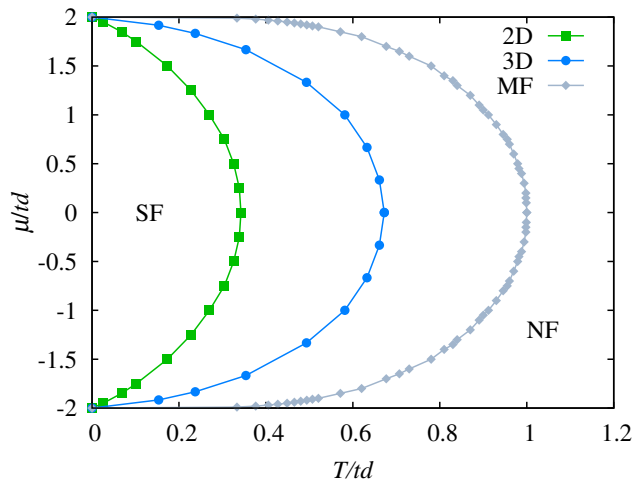}
\caption{(Color online) Finite temperature phase diagram in two and three dimensions, 
and the mean-field (MF) prediction. In all dimensions, the phase diagram contains a 
superfluid (SF) lobe surrounded by the normal fluid (NF) phase.}
\label{fig:phased}
\end{figure}

We study the Hamiltonian in Eq.~\eqref{hcb}, at finite temperature $T$, by means of the 
stochastic series expansion (SSE) quantum Monte Carlo (QMC) method with operator-loop updates 
\cite{sandvik_kurkijarvi_91,sandvik_92,dorneich_troyer_01}. The determination of the 
phase diagrams is carried out through a finite size scaling of the condensate fraction 
and/or the superfluid stiffness $\rho_s$ using periodic boundary conditions. The numerically 
exact (QMC) phase diagram in two dimensions (2D) and three dimensions, as well as the the 
mean-field predictions, are presented in Fig.~\ref{fig:phased}. The finite-temperature phase diagram 
comprises an off-diagonal long-range-ordered (ODLRO) low-temperature superfluid lobe 
(quasi-ODLRO in 2D) surrounded by a high-temperature normal phase with exponentially 
decaying correlation functions. The extend of the superfluid state is expected to be 
hindered as dimensionality is reduced because thermal and quantum fluctuations have a 
stronger effect in low-dimensional systems. Clearly, our results agree with that expectation. 
The dissimilarity between the mean-field and the exact phase diagrams makes it clear that 
both thermal and quantum fluctuations are strong and play an important role even in three 
dimensions, where mean-field approaches are generally considered to be a good approximation. 

Details on the procedure to obtain the phase boundaries are provided in the following sections. 
Such procedures are different in two and three dimensions because of the different universality 
class of the phase transition.

\section{Homogeneous systems}\label{homog}

\subsection{Two dimensions}

Our results for the two-dimensional phase diagram in Fig.~\ref{fig:phased} are based on 
the fact that the model in Eq.~\eqref{hcb} undergoes a BKT transition as a function of 
the temperature. This phase transition has been studied in great detail the context of 
the two-dimensional quantum $XY$ model in Eq.~\eqref{xym} in the absence of a magnetic field 
\cite{loh_grant_85,ding_makivic_90,ding_92,harada_kawashima_97}. Kosterlitz and Thouless
predicted that the superfluid stiffness $\rho_s$ jumps from zero to the value $(2/\pi)T_c$ 
at the critical temperature. Thus we consider measurements of the superfluid stiffness 
$\rho_s$ for different system sizes $L$ as a function of temperature. Within
the SSE method, the superfluid stiffness is computed by measuring the fluctuation of the 
winding number $W$ \cite{pollock_ceperley_87}; they are connected through the relation 
$\rho_s=\langle W^2 \rangle/2\beta$, where $\beta=1/T$ is the inverse temperature. 

\begin{figure}[!t]
\begin{center}
\includegraphics[width=0.22\textwidth,angle=-90]{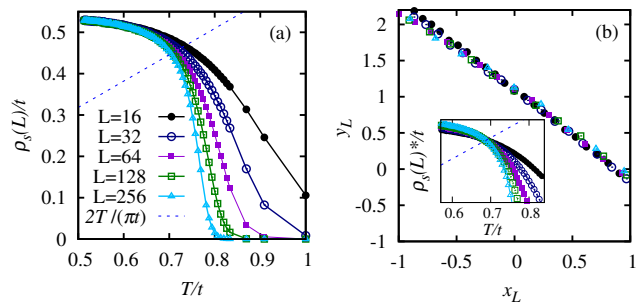}
\caption{(Color online) (a) Superfluid stiffness in 2D for $\mu=0$ and several values of $L$. The
error bars (not shown) are smaller than the point size used in the plot. 
(b) Data collapse according to the relation in Eq.~\eqref{scaling2d}. The inset in (b) shows the 
rescaled superfluid stiffness vs $T$.}
\label{fig:stiffness2d}
\end{center}
\end{figure}

Figure \ref{fig:stiffness2d}(a) shows results for the superfluid stiffness of  
2D hard-core bosons at $\mu=0$ [or, equivalently, the spin stiffness of the 2D $XY$ model in 
Eq.~\eqref{xym}] as a function of $T$ for several system sizes. The observed slow approach 
of the superfluid stiffness to the characteristic jump expected for the infinite system is 
due to strong finite-size effects at the BKT transition.  Finite-size scaling relations for 
the superfluid stiffness can be derived by integrating the Kosterlitz renormalization-group 
equations (see, for instance, 
Refs.~\onlinecite{harada_kawashima_97},\onlinecite{olsson_95},\onlinecite{weber_minhagen_88}). 
This procedure yields
\begin{eqnarray}
\frac{\rho_s\left(T,L\right)\pi}{2T}-1&=&c\coth{2c\left( \ln{L}+l_0 \right)},\quad T<T_c\nonumber\\
\frac{\rho_s\left(T_c,L\right)\pi}{2T}-1&=&\frac{1}{2\left( \ln{L}+l_0 \right)},\quad \quad T=T_c\nonumber\\
\frac{\rho_s\left(T,L\right)\pi}{2T}-1&=&c\cot{2c\left( \ln{L}+l_0 \right)},\quad T>T_c
\label{finitelstiff}
\end{eqnarray}
where $c$ measures the distance from the critical point and $l_0$ depends only weakly on temperature. 
Close to the critical point, $c\sim\sqrt{|T-T_c|}$. In the limit $2c\left( \ln{L}+l_0 \right)\ll1$, 
a scaling form for the superfluid stiffness based on Eq.~\eqref{finitelstiff} can be written as
\begin{equation}
\frac{\rho_s\left(T,L\right)\pi}{T}-2=\frac{1}
{\ln{L}+l_0}F\left[\left( \ln{L}+l_0 \right)^2\left(T-T_c \right) \right].
\label{scaling2d}
\end{equation}
From Eq.~\eqref{finitelstiff} in the limit $2c\left( \ln{L}+l_0 \right)\ll1$, $F(x)=1-\left(4/3\right)x$. 
From Eq.~\eqref{scaling2d}, one can find the scaling function $F$ and critical temperature 
$T_c$ by computing $x_L= \left(  \ln{L}+l_0 \right)^2\left(T-T_c \right)/t$ and 
$y_L=\rho_s\left(T,L\right)\pi/T-2$ based on our Monte Carlo simulations for different $L$ and 
$T$. The adjustment of the constant $l_0$ and critical temperature $T_c$, such that the data produce the 
best possible collapse, yields a numerical estimate of the scaling function $F$ and the critical 
temperature itself. The result of the determination of the scaling function $F$ is reported in 
Fig.~\ref{fig:stiffness2d}(b), where a plot of $y_L$ as a function of $x_L$ is presented. Notice 
that as expected, the value of $F$ is very close to one for $x_L=0$. Furthermore, one expects 
from  Eq.~\eqref{finitelstiff}  that a plot of the rescaled superfluid stiffness 
$\rho_s\left(T,L \right)^*=\rho_s\left(T,L\right)\left(1+\frac{1}{2\left[\ln{L}+l_0 \right]} \right)^{-1}$ 
as a function of the temperature $T$ should become system-size independent at the critical temperature 
$T_c$. This observation is confirmed in the inset of Fig.~\ref{fig:stiffness2d}(b). Remarkably, those 
curves intersect with the line $(2/\pi) T$ right at the critical temperature, in agreement with the BKT 
scenario. Our result $T_c/t=0.685 \pm0.001$ is consistent with the best value reported in 
Ref.~\onlinecite{harada_kawashima_97}, for which $T_c/t=0.6846 \pm0.0006$ \cite{note:better}. An analogous 
procedure to the one just described is carried out for different values of the chemical potential to 
complete the two-dimensional phase diagram in Fig.~\ref{fig:phased}. We should mention that 
Eq.~\eqref{finitelstiff} predicts the value of the superfluid stiffness in an infinite system at the 
critical temperature to be $\rho_s\left( T_c \right)/T_c=2/\pi$. However, in 
Ref.~\onlinecite{hasenbusch_05}, it was shown that the superfluid stiffness at the transition temperature 
is  $\rho_s\left( T_c \right)/T_c\simeq 0.63650$, which is very close to the result based on 
Eq.~\eqref{finitelstiff} ($2/\pi\simeq0.63662$). Detecting the difference is beyond the accuracy of 
the present study.

\subsubsection{Critical value from $dn_0/dT$}

\begin{figure}[!t]
\includegraphics[width=0.22\textwidth,angle=-90]{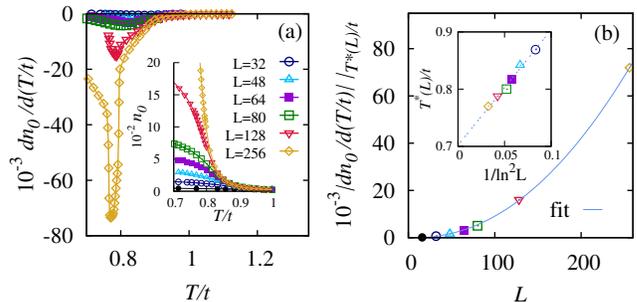}
\caption{(Color online) (a) Derivative of the zero-momentum occupation $n_0$ with respect to the 
temperature for different values of $L$. The inset shows $n_0$ vs $T$. (b) Finite-size scaling of the 
height of the negative  peak in $dn_0/dT$. The continuous line is a fit to the function  
$g(L)=a_0+a_1 L^{7/4}\ln^{3}(a_2 L)$. The inset shows the finite-size scaling of $T^{*}\left(L\right)$.} 
\label{fig:divergence}
\end{figure}

We now briefly discuss the behavior of the occupation of the zero momentum state ($n_{k=0}\equiv n_0$) 
in the critical region and address the determination of the transition temperature from it. 
In a homogeneous and infinite 3D system, BEC is identified by a macroscopic occupation of $n_0$. However, 
as mentioned before, thermal fluctuations in 2D destroy Bose-Einstein condensation. Nonetheless, 
as the superfluid transition is approached from the normal phase, $n_0$ diverges [see inset in 
Fig.~\ref{fig:divergence}(a)]. Indeed, from the Fourier transform of the one-body density matrix in the 
long-distance limit 
$\langle \hat{a}^\dagger_{i} \hat{a}^{}_{i+r} \rangle \propto r^{-1/4} \exp\left(-r/\xi\right)$, one can 
extract the behavior of $n_0$ as $T_c$ is approached,
\begin{equation}\label{n02d}
 n_0\sim \xi^{7/4}.
\end{equation}
We assume the essential singularity of the correlation length $\xi\sim e^{b/\sqrt{T-T_c}}$, where
$b$ is a chemical-potential-dependent scaling factor. From Eq.~\eqref{n02d}, it follows that not only 
does $n_0$ diverge at $T_c$, but also its derivative with respect to $T$,
\begin{equation}
\frac{dn_0}{dT} \sim -\frac{\xi^{7/4}\ln^{3}{\xi}}{b^2} 
\label{deriv_inf}
\end{equation}
In a finite system, when $T$ is close to $T_c$, the role of the correlation length is taken over 
by $L$ when  $\xi\gtrsim L$. This occurs at a characteristic temperature $T^{*}\left(L\right)$ 
given by
\begin{equation}
T^{*}\left(L\right)= T_c+ b'/\ln^2L,
\label{temp_log}
\end{equation}
where $b'$ is a nonuniversal factor related to $b$. At that temperature, the derivative in 
Eq.~\eqref{deriv_inf} scales with the system size as
\begin{equation}
\frac{dn_0}{dT}\bigg\vert_{T^{*}\left(L\right)} \sim -\frac{L^{7/4}\ln^{3}{L}}{b^2}.
\label{deriv_f}
\end{equation}

Below $T^*(L)$, $n_0$ cannot vary as fast as right above $T^*(L)$ because the 
exponential increase of the correlation length is truncated by $L$. Below $T^*(L)$, the variation of 
$n_0$ comes mainly from the temperature dependence of the anomalous exponent, which is not as strong 
as the variation due to the exponential behavior of the correlation length. Consequently, 
$dn_0/dT$ should exhibit a sharp minimum at the size-dependent temperature $T^{*}\left(L\right)$.
Moreover, in a finite system, $n_0$ cannot grow indefinitely as the temperature is lowered. With 
decreasing temperature ($T\rightarrow0$), $n_0$ must approach its (finite) $T=0$ value, which implies 
that $dn_0/dT\to 0$.

Figure~\ref{fig:divergence}(a) depicts the derivative of the $n_0$ for different system 
sizes vs $T$. The divergence of $dn_0/dT$ is apparent. A sharp minimum develops
and its location $T^{*}\left(L\right)$ approaches $T_c$ as the system size increases. This is 
expected from the finite-size relation in Eq.~\eqref{temp_log}. The scaling of the height of this minimum 
is studied in Fig.~\ref{fig:divergence}(b), where we plot the absolute value of 
$dn_0/dT|_{T^{*}\left(L\right)}$ vs $L$. The data follows the scaling relation in Eq.~\eqref{deriv_f}, 
as made evident by a fit to the function $g(L)=a_0+a_1 L^{7/4}\ln^{3}(a_2 L)$. In the inset in 
Fig.~\ref{fig:divergence}(b), we show the finite-size scaling of $T^{*}\left(L\right)$. We observe that 
$T^{*}\left(L\right)$ is consistent with the scaling relation in Eq.~\eqref{temp_log}, which we use to 
obtain the critical temperature in the thermodynamic limit. We find $T_c/t=0.701\pm0.007$. This value is 
compatible with the one found by performing the finite-size scaling of the superfluid stiffness. While 
this approach is obviously less accurate than the one discussed before for $\rho_s$, among other things 
because a numerical derivative is involved, the fact that it works extremely well is very important for 
current trapped ultracold gas experiments where the superfluid density cannot be measured. 

We note at this point that in the determination of Eq.~\eqref{deriv_f}, we have neglected multiplicative 
logarithmic corrections that affect the behavior of the zero-momentum occupation and thus its derivative 
with respect to the temperature ~\cite{amit_grinstein_80,kenna_irving_95}. In fact, the exponent of the 
logarithm in Eq.~\eqref{deriv_f} gets modified to
\begin{equation}
\frac{dn_0}{dT}\bigg\vert_{T^{*}\left(L\right)} \sim -\frac{L^{7/4}\ln^{3-2r}{L}}{b^2},
\label{deriv_f_modified}
\end{equation}
with $r=-1/16$ (Ref.~\onlinecite{kenna_irving_95}). However, this correction does not affect 
the determination of the critical temperature, which is based on the location of the position of the peak 
in the numerical derivative and the scaling relation in Eq.~\eqref{temp_log}. Furthermore, the correction 
to the exponent of the logarithm is very small and, at least within the precision of our simulations, 
its effect is hardly detectable.

\subsection{Three dimensions}

\begin{figure}[!t]
\includegraphics[width=0.22\textwidth,angle=-90]{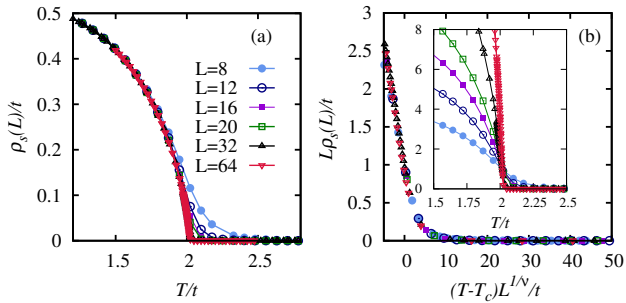}
\caption{(Color online) (a) Superfluid stiffness of the 3D system for $\mu=0$ and several 
values of $L$. (b) Data collapse according to the relation in Eq.~\eqref{stiff3dscaling}. 
The inset shows the rescaled superfluid stiffness as a function of $T$.}
\label{fig:stiffness3d}
\end{figure}

In order to determine the 3D phase diagram, we follow the same procedure as in 2D. In 3D, however, 
the superfluid to normal transition belongs to the 3D $XY$ universality class. This transition, 
for the model in Eq.~\eqref{hcb}, has also been studied using QMC simulations in the past. $T_c$ 
for BEC was evaluated as a function of the density in Ref.~\onlinecite{pedersen_schneider_96}. 
The onset of magnetization as a function of the magnetic field (or, in the bosonic language, the density 
as a function of the chemical potential) was investigated in Ref.~\onlinecite{kawashima_04}. Furthermore, 
the fate of the superfluid phase under the effect of an additional ring-exchange term was studied in 
Ref.~\onlinecite{melko_scalapino_05}. Here, we determine the full phase diagram (shown in 
Fig.~\ref{fig:phased}) as a function of the temperature and the chemical potential. We begin 
by considering measurements of the superfluid stiffness. In $d>2$ dimensions, as the critical 
temperature is approached, the superfluid stiffness vanishes continuously as \cite{fisher_jasnow_73} 
\begin{equation}
\rho_s \sim |T_c -T|^{\left(d-2\right)\nu },
\end{equation}
where the exponent $\nu$ determines how the correlation length diverges when approaching the critical 
temperature, i.e.,
\begin{equation}
 \xi \sim |T-T_c|^{-\nu}.  
\label{corrle}
\end{equation}
As a result, at the critical temperature, the superfluid stiffness scales with the linear size of the 
system as $\rho_s \sim L^{2-d}$. This, in turn, allows one to write the scaling hypothesis for the 
superfluid stiffness as a function of the system size and the temperature as 
\begin{equation}
\rho_s L^{d-2} = F\left( |T-T_c|L^{1/\nu} \right),
\label{stiff3dscaling}
\end{equation}
which we utilize to determine the critical temperature. In Fig.~\ref{fig:stiffness3d}(a), we show 
results for the superfluid stiffness in a 3D lattice vs $T$ for different system sizes. 

We numerically extract the scaling function $F$ by studying the rescaled superfluid stiffness 
[left-hand side in Eq.~\eqref{stiff3dscaling}] vs the rescaled temperature $(T-T_c)L^{1/\nu}$.
Classical Monte Carlo simulations yield the correlation length exponent $\nu= 0.6717\pm0.0001$
\cite{campostrini_vicari_06}, and $\nu= 0.6717\pm0.0003$ \cite{burovski_svistunov_06}, which we use to 
produce the  collapse presented in Fig.~\ref{fig:stiffness3d}(b). With $\nu$ at hand, it is enough to 
fix $T_c$ such that the best collapse of the data is achieved. Furthermore, the inset shows the 
rescaled superfluid stiffness as a function of temperature, which becomes system-size independent 
at the critical temperature, as implied by the scaling hypothesis in Eq.~\eqref{stiff3dscaling}. Our 
best estimation of the critical temperature for $\mu=0$ is $T_c/t=2.0169\pm0.0005$ (to be compared with 
$T_c/t=1.94$ from Ref.~\onlinecite{pedersen_schneider_96} and more recently with $T_c/t=2.016\pm0.004$ 
from Ref.~\onlinecite{laflorencie_12}). We perform a similar analysis for different values of the 
chemical potential to complete the three-dimensional phase diagram in Fig.~\ref{fig:phased}.

Additionally, since the superfluid to normal phase transition in our model in 3D is 
accompanied by the emergence of true long-range order, one can study the transition by computing 
the condensate fraction $f_0$ associated with the appearance of BEC. Following 
Penrose and Onsager \cite{penrose56}, the condensate fraction is defined as the ratio of the largest 
eigenvalue of the one-body density matrix to the total number of particles $N_b$. For the system under 
consideration, condensation occurs to the zero-momentum state due to translational invariance, thus the 
condensate fraction is  $f_0=n_0/N_b$. The behavior of $n_0$ can 
be obtained from the Fourier transform of the one-body density matrix in the long-distance limit, 
which in 3D is given by
\begin{equation}
\langle \hat{a}^\dagger_{i} \hat{a}^{}_{i+r} \rangle \propto r^{-\left(1+\eta\right)} 
\exp\left(-r/\xi\right).
\label{obdm_3d}
\end{equation}
Here, $\eta$ is the correlation function exponent, also known as the anomalous scaling dimension. On approach to
$T_c$, $n_0$ diverges with the correlation length as 
\cite{pollet_svistunov_10}
\begin{equation}
n_0 \sim \xi^{2-\eta}.
\label{n03d}
\end{equation}
In a finite system, this relation implies that the condensate fraction vanishes at the critical point as 
$f_0 \sim L^{-(1+\eta)}$, which we adopt to formulate the following scaling hypothesis for the condensate 
fraction
\begin{equation}
f_0 L^{1+\eta}= F\left(|T-T_c|L^{1/\nu} \right).
\label{n03dscalinscalingg}
\end{equation}

\begin{figure}[!t]
\includegraphics[width=0.22\textwidth,angle=-90]{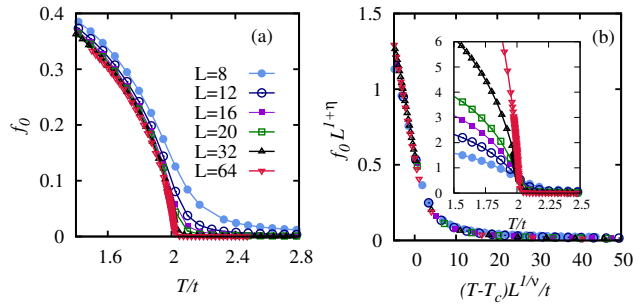}
\caption{(Color online) (a) Condensate fraction in 3D for $\mu=0$ and several values of $L$. (b) Data 
collapse according to the relation in Eq.~\eqref{n03d}. The inset shows the rescaled condensate fraction 
as a function of $T$.}
\label{fig:condensate3d}
\end{figure}

In the determination of $T_c$ through the scaling relation in Eq.~\eqref{n03dscalinscalingg}, we 
use the value $\eta=0.0381\pm0.0002$ \cite{campostrini_vicari_06}. The results are summarized in 
Fig.~\ref{fig:condensate3d}, where a plot of the condensate fraction vs $T$ is shown 
in panel (a). In Fig.~\ref{fig:condensate3d}(b), the data collapse of the rescaled condensate 
fraction $f_0 L^{1+\eta}$ vs the rescaled temperature is apparent. Furthermore, in the inset, one can 
observe that curves of the rescaled condensate fraction vs $T$ become system-size independent at 
$T_c$, as implied in Eq.~\eqref{n03dscalinscalingg}. This procedure results in a $T_c/t=2.0167\pm0.0005$ 
for $\mu=0$, which is in remarkably good agreement with our previous estimate using the superfluid stiffness.

\subsubsection{Critical value from $dn_0/dT$}

\begin{figure}
\includegraphics[width=0.22\textwidth,angle=-90]{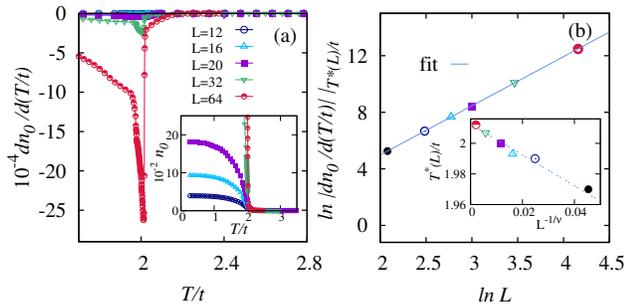}
\caption{(Color online) (a) $dn_0/dT$ for different values of $L$. The inset shows $n_0$ vs $T$. 
(b) Finite-size scaling of the height of the negative peak in $dn_0/dT$. The continuous line is a fit to 
the function  $g(\ln L)=a_0+a_1 \ln L$. The inset shows the finite-size scaling of $T^{*}\left(L\right)$. }
\label{fig:divergence3d}
\end{figure}

Similarly to the 2D case, $dn_0/dT$ diverges in the vicinity  of the superfluid to normal phase 
transition. It diverges with the correlation length as
\begin{equation}
\frac{dn_0}{dT} \sim -\xi^{2-\eta+1/\nu}.
\label{deriv_3d}
\end{equation}
Also, as in 2D, in a finite 3D system at a temperature $T^{*}\left(L\right)$ close to $T_c$, 
the role of the correlation length is taken over by $L$ when $\xi\gtrsim L$. The characteristic 
temperature $T^{*}\left(L\right)$ is given by
\begin{equation}
T^{*}\left(L\right)= T_c+ c'/L^{1/\nu},
\label{temp_asterisk_3d}
\end{equation}
where $c'$ is a non-universal factor. At $T^{*}\left(L\right)$, $dn_0/dT$ scales with the system 
size as
\begin{equation}
\frac{dn_0}{dT}\bigg\vert_{T^{*}\left(L\right)} \sim -L^{2-\eta+1/\nu}.
\label{deriv_finite_3d}
\end{equation}
Furthermore, in a finite system, $dn_0/dT$ reaches its minimum value at $T=T^{*}\left(L\right)$ 
because the divergence of the correlation length can no longer be sustained. This is expected 
from the behavior of $n_0$ vs $T$, shown in the inset in Fig.~\ref{fig:divergence3d}(a), where $n_0$ is 
first seen to increase as the temperature is lowered and then to saturate as $T\rightarrow0$. The changes 
observed $dn_0/dT$ in that low temperature regime originate in the smooth dependence of the correlation 
function exponent on the temperature, as opposed to the fast change produced by the strong divergence of 
the correlation length. Hence, once again, $dn_0/dT$ exhibits a sharp minimum at the size-dependent 
temperature $T^{*}\left(L\right)$ in Eq.~\eqref{temp_asterisk_3d} and then goes to zero.

In Fig.~\ref{fig:divergence3d}(a), we display results for $dn_0/dT$ vs $T$ for different system sizes. 
The divergence in the derivative, anticipated by Eqs.~\eqref{deriv_3d} and \eqref{deriv_finite_3d}, 
is confirmed by the presence of sharp minima that grow with system size. The finite-size scaling of the 
height of the sharp minimum in Eq.~\eqref{deriv_finite_3d} is presented in Fig.~\ref{fig:divergence3d}(b), 
where we plot the logarithm of the maximum height of $|dn_0/dT|$ vs $\ln L$. According to 
Eq.~\eqref{deriv_finite_3d}, such a plot should turn into a straight line with a slope given by 
$m=2-\eta+1/\nu$. A fit of our data to the function $g(\ln L)=a_0+a_1 \ln L$, yields  $a_1=3.47 \pm 0.01$. 
The scaling relation given by Eq.~\eqref{deriv_finite_3d} is thus confirmed as our value of $a_1$ is 
compatible with the exponents from Ref.~\onlinecite{campostrini_vicari_06}, which yield $m=3.450$. The 
size dependence of the position of the peaks anticipated in Eq.~\eqref{temp_asterisk_3d} is verified in 
the inset of Fig.~\ref{fig:divergence3d}(b). Within this procedure, we find that the critical temperature 
in the thermodynamic limit is $T_c/t=2.012\pm 0.002$, which is in relatively good agreement with the one 
obtained through the finite-size scaling of both the superfluid stiffness and the condensate fraction. 

We conclude this section by mentioning that in determining the critical temperature, we have used 
the leading scaling forms and subleading corrections to scaling have been 
neglected. For the 3D $XY$ universality class, such corrections have been reviewed in 
Ref.~\onlinecite{pelissetto_vicari_02}. We note that in our calculations, there is an excellent collapse 
of the data, which suggests that the effects of the subleading corrections to scaling are small. 
Furthermore, the most accurate results obtained for $T_c$ follow from completely independent measurements, 
i.e., the superfluid and condensate fractions. They agree within the error bars, which further 
supports the relevance of the scaling relations used.

\subsection{Mean field}

To gain an understanding of the effects of quantum fluctuations in our systems, we have also calculated 
the mean-field phase diagram for this model. We utilize the standard decoupling of the kinetic energy term 
in the Hamiltonian in Eq.~\eqref{hcb} \cite{sheshadri_Krishnamurthy_93}
\begin{equation}\label{decop}
{\hat a}^\dagger_i {\hat a}_j \simeq {\hat a}^\dagger_i  \Phi_j+{\hat a}_j \Phi_i^* - \Phi_i^* \Phi_j,
\end{equation}
where $ \Phi_i=\langle {\hat a}_i \rangle $ is the condensate order parameter, to be determined 
self-consistently. The angle brackets denote the usual thermal average. The above mean-field decoupling 
allows one to write a mean-field Hamiltonian for Eq.~\eqref{hcb} as
\begin{align}\label{meanfbose1}
\hat{{\cal H}}_{\textrm{MF}}=- t\sum_{\langle i,j \rangle} 
\left( {\hat a}^\dagger_i \Phi_j + \Phi_i^* {\hat a}_j-\Phi_i^*\Phi_j \right) 
+ \textrm{H.c.} - \sum_i  \mu_i {\hat n}_i.
\end{align}

For homogeneous systems, i.e., $V_0=0$, Eq.~\eqref{meanfbose1} can be recast in the following manner,
\begin{align}\label{meanfbose2}
\hat{{h}}_{\textrm{MF}}=- 2dt\,\Phi \left( {\hat a}^\dagger + {\hat a}\right)
-  \mu {\hat n}, 
\end{align}
where $\hat{{h}}_{\textrm{MF}}$ is the mean-field Hamiltonian per lattice site. Note that in this case 
the superfluid order parameter can be taken to be real. The corresponding partition function at finite 
inverse temperature $\beta$ is
\begin{equation}\label{partition}
Z=2 e^{-\beta\frac{\mu}{2}}\cosh{\beta \sqrt{\frac{\mu^2}{4}+\left(2dt\,\Phi\right)^{2}}}.
\end{equation}

A self-consistency condition for the superfluid order parameter can be derived by noting that
\begin{equation}\label{partition_self}
\frac{d Z}{d\Phi}= 4\beta dt \langle {\hat a} \rangle Z.
\end{equation}
Using the relation \eqref{partition_self}, we arrive at the equation that determines the order parameter 
$\Phi$,
\begin{equation}\label{selfconsistency}
 \sqrt{\frac{\mu^2}{4}+\left(2dt\,\Phi\right)^{2}}= dt \tanh { \beta \sqrt{\frac{\mu^2}{4}+
\left(2dt\,\Phi\right)^{2}}}, 
\end{equation}
which is valid whenever $\Phi>0$. We solve Eq.~\eqref{selfconsistency} numerically and determine the 
superfluid region, $\Phi>0$, as a function of the temperature and the chemical potential. The phase 
boundaries are determined as the values of $\mu$ and $T$ for which $\Phi\to 0$. For $\mu=0$, 
Eq.~\eqref{selfconsistency} reduces to 
\begin{equation}\label{ising}
2\Phi=\tanh{\beta\, 2dt\, \Phi}, 
\end{equation}
which is the equation that determines the mean-field magnetization of the Ising model in 
the absence of a magnetic field. The critical temperature is, of course $T_c/td=1$, which is quite 
different from the results of our quantum Monte Carlo simulations in two and three dimensions.

\section{Trapped systems}\label{trapped}

In experiments involving ultracold atoms, an additional trapping potential is necessary to contain the gas.
While a qualitative (and sometimes a reasonably good quantitative) description of the trapped system can be 
obtained within the local density approximation (LDA) from the properties of the homogeneous system,
this approximation may breakdown in regimes of interest. In particular, the latter occurs at criticality, 
where the correlation length diverges and deviations from the LDA description can be large 
\cite{pollet_svistunov_10}. Furthermore, as we explain below, in trapped 2D systems care needs 
to be taken with the application of the Mermin-Wagner-Hohenberg theorem. Therefore, we focus our attention 
on those two aspects, namely, the possibility to have BEC the in the presence of an additional external confining 
potential in 2D, and the study of criticality in 2D and 3D.

\subsection{Absence of BEC in interacting 2D systems}

We mentioned in the Sec.~\ref{intro} that homogeneous 2D systems are special because thermal fluctuations 
destroy any order at finite temperature. However, harmonically confined non-interacting bosons can 
undergo BEC at finite temperature \cite{dalfovo_giorgini_review_99}. In this case, the arguments 
by Mermin {\it et al.} are not violated because condensation does not occur to the zero-momentum 
state, but to a single-particle eigenstate of the trapped system. One can then wonder whether 
finite-temperature BEC persists in the presence of interactions. By following analogous arguments to those 
in Ref.~\onlinecite{mullin_97}, we show below that interactions do preclude the formation of a condensate 
in the Bose-Hubbard model in the presence of the trap. This is so because there is a close  connection 
between the formation of a condensate and the macroscopic population of the zero-momentum occupation, 
which is forbidden in 2D at finite temperature. 

Generally speaking, the emergence of BEC is established through the evaluation of the condensate fraction 
$f_0$, which is defined as the ratio of the largest eigenvalue of the one-body density matrix $n_M$ to 
the total number of particles $N_b$,
\begin{equation}\label{bec_generic}
f_0=\frac{n_M}{N_b}.
\end{equation}
If after taking the appropriate thermodynamic limit $f_0$ remains finite, then the system exhibits BEC. 
Otherwise, if it becomes zero, there is no condensation \cite{penrose56}.

Alternative forms of the criteria expressed through Eq.~\eqref{bec_generic} can be  useful when the system 
is not spatially  uniform; they are based on the following inequality \cite{penrose56}:
\begin{equation}\label{inequ0}
n_M^2\leq \sum_{a}n_a^2\leq n_M \sum_{a}n_a=n_M N_b,
\end{equation}
where $n_a$ are the eigenvalues of the one-body density matrix $\rho_{ij}$. We define the quantity
\begin{equation}\label{bec_crit}
A_2=N_b^{-2} \sum_{i,j} |\rho_{ij}|^2,
\end{equation}
which is just a lattice version of its analogous quantity defined on the continuum in 
Ref.~\onlinecite{penrose56}. It follows from Eqs.~\eqref{inequ0} and ~\eqref{bec_crit} that
\begin{equation}\label{inequ}
f_0^2\leq A_2\leq f_0. 
\end{equation}
Therefore, if $A_2$ remains finite in the thermodynamic limit, then the system exhibits BEC. 
A further criterion can be defined and it depends on the quantity
\begin{equation}\label{bec_crit2}
A_1=\left(N_b L^d\right)^{-1} \sum_{i,j} |\rho_{ij}|.
\end{equation}
Notice that $\left(A_1 N_b/L^d\right)^2$ is the square of the mean value of the function $|\rho_{ij}|$, 
while $A_2\left(N_b/L^d\right)^2$ is the mean value of $ |\rho_{ij}|^2$. Since the variance of the 
function $|\rho_{ij}|$ is either positive or zero, it follows that 
\begin{equation}\label{ineq2}
A_1^2\leq A_2.
\end{equation}
Now, since $\rho_{ij}$ is a positive-semidefinite Hermitian matrix, its elements satisfy \cite{penrose56,horn_85}
\begin{equation}\label{positiveherm}
|\rho_{ij}|\leq \sqrt{\rho_{ii}\,\rho_{jj}}\leq \frac{1}{2}\left(\rho_{ii}+\rho_{jj}\right) \leq \alpha N_b/L^d,
\end{equation}
where $\alpha N_b/L^d$ is an upper bound of the local density $\rho_{ii}$. By summing over $i$ and $j$ in  
Eq.~\eqref{positiveherm} and the square of it, we find a lower bound for $A_1$,
\begin{equation}\label{ineq3}
A_2\leq \alpha A_1. 
\end{equation}

As long as the local density $\rho_{ii}$ remains finite  throughout the whole system, $\alpha$ can be taken 
to be finite and independent of $N_b/L^d$. This, in turn, implies that if $A_1>0$, then BEC takes place; 
otherwise if $A_1=0$, then no BEC occurs \cite{penrose56}. Notice that if $\rho_{ij}\geq0$, then $A_1$ 
coincides with the ratio of the zero-momentum occupation to the total number of particles, i.e., the 
fraction of particles in the system that condenses to the zero-momentum state. Since in two dimensions 
$n_0/N_b$ vanishes because of the Mermin {\it et al.} theorem, then $A_1$ is zero too. In the specific 
case of the Bose-Hubbard model in the presence of an inhomogeneous potential in thermal equilibrium, we 
have that $\rho_{ij}\geq0$. Furthermore, the density is finite everywhere across the system because of 
the on-site interaction, implying that $A_1=0$. 

Hence,  even in the presence of the trap, there is no condensation in the 2D Bose-Hubbard model at 
finite $T$. Note that this argument does not preclude condensation in the non-interacting limit, where 
the density can diverge at the minimum of the inhomogeneous potential in the thermodynamic limit and BEC 
can indeed occur to the lowest single-particle eigenstate, but not to the zero-momentum state. Moreover, 
the criteria above implies that for the Bose-Hubbard model in $d>2$ in thermal equilibrium, condensation 
to any state has to be accompanied by condensation to the zero-momentum state.

In our proof, we have stated that for the Bose-Hubbard model in thermal equilibrium $\rho_{ij}\geq0$ 
holds. We now present two independent arguments for why $\rho_{ij}\geq0$. The first one is based on 
the fact that the matrix elements of the von Neumann's statistical operator in the position 
representation are strictly positive \cite{campbell_senger_84}. Since the one-body density matrix 
corresponds to a partial trace of the von Neumann's statistical operator \cite{penrose56}. it follows 
that its elements are positive too. A rather technical, but yet rigorous, argument is based on the 
series expansion representation of the one-body density matrix that we used in our Monte Carlo 
implementation. Within this representation, the measurements of the one-body density matrix are based 
on the extension of the configuration space where these off-diagonal quantities are well 
defined \cite{dorneich_troyer_01}. In such extended space, the one-body density matrix is represented 
as the sum of strictly positive matrix elements (hence $\rho_{ij}\geq0$) which are, in turn, 
efficiently sampled during the construction of the loop operators in the directed-loop update 
algorithm \cite{alet_troyer_05}.

\subsection{Two dimensions}

\subsubsection{Local compressibility}

Exactly as in the homogeneous system, even though there is no condensation in 2D, a superfluid phase is 
expected in the trapped system at low temperatures. Because of the inhomogeneity introduced by the 
confining potential, the coexistence of space separated normal and superfluid domains can occur at 
intermediate temperature. In that case, there must be a region in the trap where superfluidlike domains 
transition into normal ones. Within the LDA, this  region is such that the local chemical potential 
$\mu_i$ coincides with the critical $\mu$ of the bulk system for the normal-to-superfluid phase transition. 

Based on this idea, Zhou and collaborators proposed a method to identify the phase boundaries 
of the homogeneous system from a high-resolution scan of the local density $\rho(r)$ across 
the confined system \cite{zhou_trivedi_09}. This method requires the determination of the local 
compressibility defined as
\begin{equation}
\kappa_{\textrm{diff}}\left(r\right)=-\frac{1}{2V_0 r}\frac{d \rho \left(r\right)}{dr},
\end{equation}
and relies on the expectation that the local density profile $\rho(r)$, as well as the local 
compressibility $\kappa_{\textrm{diff}}\left(r\right)$, can be well approximated by their bulk values 
through the LDA. The existence of sharp features in the local compressibility at specific locations 
in the trap is then associated with phase transitions occurring in the homogeneous system as a function 
of the chemical potential. This method is expected to be accurate in the limit of very shallow 
traps where the contribution from density gradients due to the trapping potential are 
small \cite{zhou_trivedi_10}. 

\begin{figure}[!b]
\includegraphics[width=0.2\textwidth,angle=-90]{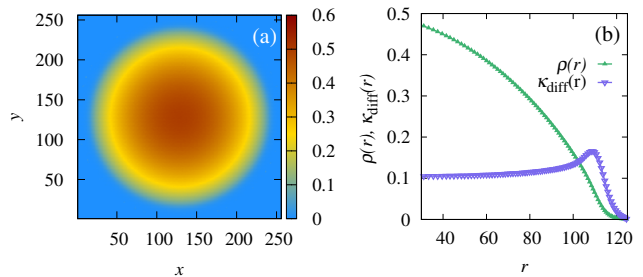}
\caption{(Color online) (a) Two-dimensional density at $T/t=0.2012$ and a trapping potential 
$V_0/t=0.0003$, for $\mu=0$ in the center of the trap. (b) The corresponding density 
profile $\rho(r)$, as well as the local compressibility $\kappa_{\textrm{diff}}\left(r\right)$, 
as a function of the distance from the center of the trap $r$. All distances $x$, $y$, and $r$ are 
measured in units of $a$ while the local compressibility is measured in units of $1/t$}
\label{fig:locdensity}
\end{figure}

In Fig.~\ref{fig:locdensity}, we present QMC results for the density profile of a 2D trapped system, 
as well as the local compressibility, as a function of the distance from the center of the trap. 
The expected sharp features in the local compressibility due to critical fluctuations are smoothed by
finite-size effects. They are replaced by a rounded maximum, which can be associated with the 
superfluid to normal transition \cite{zhou_trivedi_10}.  The location of the maximum $r_c$ is connected 
to the critical chemical potential through $\mu_c=\mu-V_0 r_c^2$. For the case in Fig.~\ref{fig:locdensity}, 
we get $\mu_c/t=-3.57 \pm 0.03$. This value is to be contrasted with $\mu_c/t=-3.5$, which we obtained 
in the homogeneous system calculations. As $T$ increases, however, the agreement between the estimates of 
the critical chemical potential based on the local compressibility and the results of the homogeneous system 
worsens. For instance, for $T/t=0.4562$, we find that $\mu_c/t=2.99\pm0.04$, as opposed to the homogeneous 
system result where $\mu_c/t=2.5$. This occurs presumably because, closer to the tip of the superfluid 
lobe, critical fluctuations are stronger, and thus larger violations of the LDA are expected.
 
\subsubsection{Momentum distribution function}

Another quantity that can be measured in experiments with ultracold atoms is the momentum distribution 
function. At fixed chemical potential ($\mu\leq0$), when lowering $T$, the normal-to-superfluid crossover 
in the trapped system proceeds via the creation and growth of a superfluid domain in the center of the 
trap. (The rate of growth of the superfluid domain will depend on the functional form and strength of 
the confining potential.) Hence, the zero-momentum state becomes increasingly populated. As follows from 
the discussion for finite homogeneous systems, it is expected that as $T$ decreases and approaches $T_c$ 
for the normal-to-superfluid transition in the center of the trap, the rate of growth of $n_0$ will 
increase. Below $T_c$, on the other hand, $dn_0/dT$ will eventually decrease because of the finite extend 
of the system imposed by the confining potential. If $T$ is lowered well below $T_c$, then almost the entire 
system will become superfluid and the observables will saturate their (finite) zero-temperature values.
 
\begin{figure}[!t]
\includegraphics[width=0.225\textwidth,angle=-90]{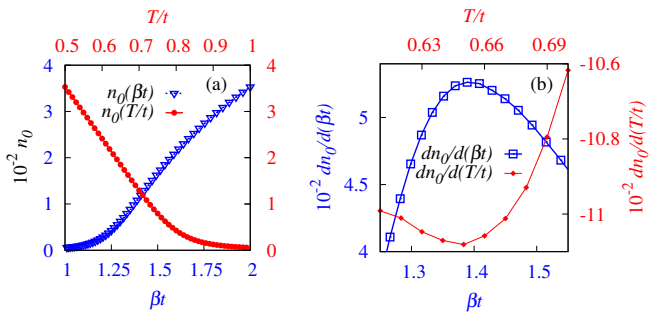}
\caption{(Color online) (a) $n_0$ as a function of $T$ and $\beta$ in a trapped 2D system with 
$V_0/t=0.00125$, $\mu=0$ in the center of the trap, and $L=128$. (b) Derivatives of $n_0$ with respect 
to $\beta$  and with respect to $T$.}
\label{fig:derivtrapped}
\end{figure}

Hence, just as in the homogeneous case, one can attempt to estimate $T_c$ for the superfluid to normal 
phase transition for the density in the center of the trap by measuring the temperature at which the 
rate of change of $n_0$ is extremal. This approach provides an accurate estimate for the homogeneous 
system and it is expected to be accurate in confined systems with shallow trapping potentials. 
Figure \ref{fig:derivtrapped}(a) depicts the evolution of $n_0$ vs $T$ as well as the inverse temperature 
$\beta$ of a harmonically confined 2D system with $V_0/t=0.0015$ ($L=128$) and $\mu=0$ in the center of 
the trap. In Fig.~\ref{fig:derivtrapped}(b), we show $dn_0/dT$ which, as expected, exhibits a minimum 
located at $T/t=0.66\pm0.02$. This temperature is compatible with the value of $T_c/t$ obtained for the 
homogeneous case where, after a finite-size scaling, we obtained $T_c/t=0.685 \pm0.001$. Our estimate 
derived from the study of a single trapped system is about $4\%$ off the value of the homogeneous system.

One can perform the same analysis based on measurements of $n_0$, but now as a function of the inverse 
temperature $\beta$. In that case, one expects a maximum in the derivative $d n_0/d\beta$ instead of a 
minimum. In general, for finite and not very large systems, the position of such maximum $\beta_c$ will 
not coincide with $1/T_c$ obtained from the minimum of $d n_0/d T$. Overall, we find that for the 
system sizes available to our QMC simulations, the analysis based on $d n_0/d\beta$ provides more 
accurate estimates of the critical temperature than the one based on $d n_0/d T$. Furthermore, the 
maximum found in $d n_0/d\beta$ is consistently sharper and better defined with respect to the minimum 
found for $d n_0/d T$ which instead is shallower and broader, and thus harder to detect and numerically 
less reliable. 

Based on measurements of $d n_0/d\beta$ presented in Fig. \ref{fig:derivtrapped}(b) on 
the same system with $V_0/t=0.0015$ ($L=128$), $\mu=0$, we find $T_c/t=0.72\pm0.02$, which is also very 
close  to the critical temperature of the homogeneous system. When the maximum is sharply defined, 
in the limit of very shallow traps with large numbers of bosons, the two approaches are expected to 
coincide (i.e., their difference is due to finite-size effects). As a matter of fact, for the homogeneous 
2D and 3D systems in Sec.~\ref{homog}, where the minima of $dn_0/dT$ are sharp, we find that the 
analysis using $dn_0/dT$ and $d n_0/d\beta$ yields essentially the same results for $T_c$. In the 
Appendix, we provide an analytic understanding of this in terms of a simple 
function. Therefore, for the determination of the phase diagram based on measurements in harmonically 
confined systems, we consider only measurements based on $d n_0/d\beta$.

\begin{figure}[!t]
\includegraphics[width=0.32\textwidth,angle=-90]{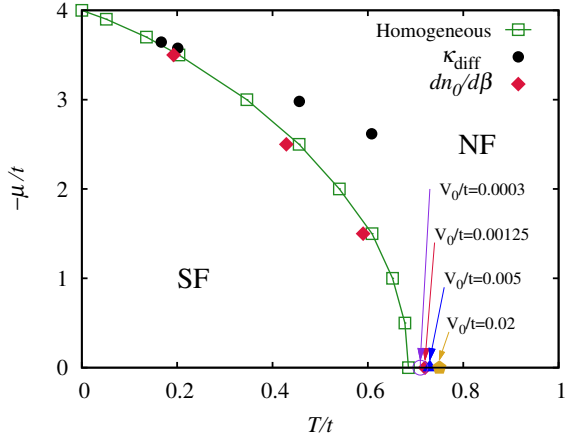}
\caption{(Color online) Estimate of the critical points based on the local compressibility (black dots 
based on a system with $L=256$) as well as the derivative of the zero-momentum occupation with respect 
to $\beta$ (red diamonds based on a system  with $L=128$). At the tip of the superfluid lobe we include 
further results for different system sizes and trap strengths (yellow pentagon $L=32$, blue triangle 
$L=64$, violet empty circle $L=256$). The phase diagram of the homogeneous system is also shown.}
\label{fig:comparisons}
\end{figure}

In Fig.~\ref{fig:comparisons}, we summarize our results for the determination of the critical parameters 
with the local compressibility as well as with the derivative of the zero-momentum occupation with respect
 to $\beta$, and contrast them with the phase diagram of the homogeneous system. Clearly, all methods 
work well for large values of $\mu/t$ and small values of $T_c/t$ (equivalent to approaching the continuum
limit in a lattice system). Close to the tip of the superfluid region, the method based on $n_0$ performs 
much better than the one based on $\kappa_{\textrm{diff}}\left(r\right)$.

At the tip of the superfluid lobe, where the size effects are expected to be the strongest, we observe 
that as the size of the system is increased (or the strength of the trap is decreased), keeping constant 
the chemical potential in the center of the trap, the estimate of the critical temperature decreases 
approaching the result in homogeneous systems.

\subsection{Three dimensions}

We now turn our attention to the study of criticality in 3D trapped systems. We make use of the same 
ideas developed for 2D system to extract the critical parameters, i.e., measurements based on the 
zero-momentum occupation as well as on the local compressibility. 

Additionally, in 3D, we can utilize a method that is based on the analysis of the shape 
of the central peak for the momentum distribution. With it, one can construct a quantity that exhibits a 
minimum at the critical point \cite{pollet_svistunov_10}. The idea behind this method is that close to 
criticality, the momentum distribution develops a bimodal structure whose evolution as a function of 
temperature contains information about the formation of a superfluid region in the center of the 
trap. At $T_c$, when a superfluid domain begins to form, the major contribution to the occupation of 
the zero-momentum state comes from regions that are not critical, i.e., from regions that are far away 
from the center of the trap. However, the derivatives of the momentum distribution $d^{m} n_k/dk^{m}$ are 
critical, in the sense that they can be understood in terms of an LDA integral that diverges at the 
center of the trap, where the system is critical. Based on that idea, the following quantity was 
devised in order to extract the critical temperature \cite{pollet_svistunov_10}:
\begin{equation}\label{qoft}
Q\left(T\right)= \left( n_{0} - n_{k_{\textrm{max}}} \right) (k_{\textrm{max}})^s,
\end{equation}
where $k_{\textrm{max}}$ is the momentum at which $|dn_k/dk|$ is maximum and the exponent $s>2-\eta$. In 
Ref.~\onlinecite{pollet_svistunov_10}, it was shown that $Q\left(T\right)$ should exhibit a minimum at 
the critical temperature $T_c$.

\begin{figure}[!t]
\includegraphics[width=0.24\textwidth,angle=-90]{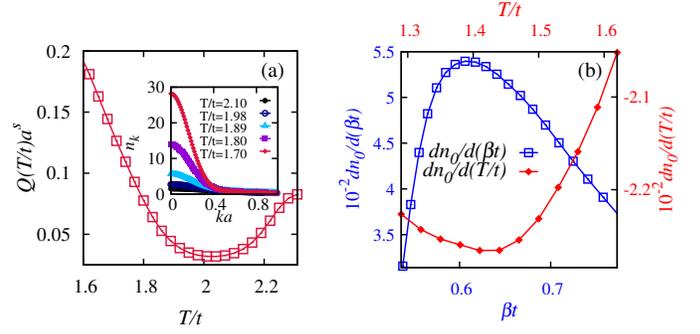}
\caption{(Color online) (a) The quantity $Q\left(T\right)$ as a function of temperature
extracted from the momentum distributions shown in the inset. The exponent in Eq.~\eqref{qoft} has 
been set to $s=3$. (b)  Derivatives of $n_0$ with respect to $\beta$ and with respect to $T$.
 The three-dimensional system is prepared with $V_0/t=0.04$, $\mu=0$, and $L=32$.}
\label{fig:qt}
\end{figure}

We plot $Q\left(T\right)$ vs $T$ in Fig.~\ref{fig:qt}(a). $Q\left(T\right)$ exhibits a minimum at 
$T_c/t=2.04 \pm0.03$. In the inset, we show the evolution of the momentum distribution function as the 
temperature of the system is reduced. This result is compatible with the critical temperature found 
for the homogeneous system $T_c/t=2.0169\pm0.0005$. In principle, similar ideas as the ones presented 
in Ref.~\onlinecite{pollet_svistunov_10} could be used to devise a quantity $Q\left(T\right)$ to 
locate the critical parameters in 2D. In that case, however, the structure of the momentum distribution 
is different because the transition is in another universality class. As a result, the LDA integrals 
for the central peak and the derivatives of the momentum distribution get substantially 
modified. We find that both the central peak and the derivatives of $n_k$ are critical in 2D because 
the LDA integrals of those quantities diverge in the center of the trap where the system is critical. 
Hence, one cannot define a $Q\left(T\right)$, as done in 3D, that will exhibit a minimum at $T_c$. 

In Fig.~\ref{fig:qt}(b), we also display results obtained for $d n_0/d\beta$ ($d n_0/dT$) in the same 
system. The temperature at which the maximum (minimum) occurs for those quantities exhibits a larger 
deviation from $T_c$, from the homogeneous case, than $Q(T)$. However, with increasing system size, 
we find that the maxima of $d n_0/d\beta$ (minima of $d n_0/dT$) slowly approach the homogeneous 
result. In experiments where the system sizes are much larger than the ones studies here, we expect 
that $d n_0/dT$ and $d n_0/d\beta$ will both produce accurate results for $T_c$.

\begin{figure}
\includegraphics[width=0.32\textwidth,angle=-90]{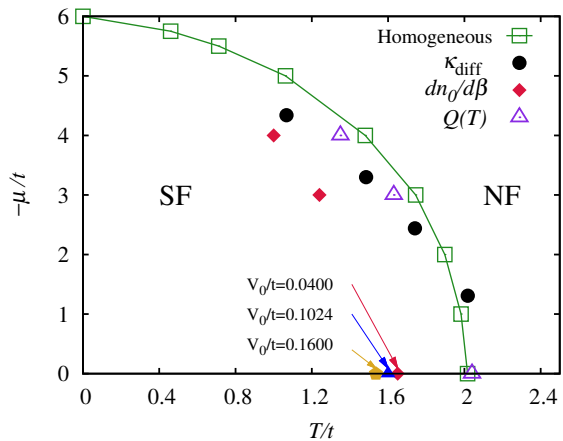}
\caption{(Color online) Estimates of the critical points based on the local compressibility (black 
dots; based on a system with $L=64$) as well as the derivatives of the zero-momentum occupation 
with respect to $\beta$ (red diamonds based on a system with $L=32$). At the tip of the superfluid 
lobe, we include further results for different system sizes and trap strengths (yellow pentagon: 
$L=16$; blue triangle: $L=20$). The estimates based on Eq.~\eqref{qoft} are shown by purple 
empty triangles. The phase diagram of the homogeneous system is also drawn with green empty squares. }
\label{fig:comp3d}
\end{figure}

In Fig.~\ref{fig:comp3d}, we present a summary of our estimates of the critical parameters based on 
the local compressibility, the derivatives of $n_0$ with respect to $\beta$, and Eq.~\eqref{qoft}. 
The method based on $Q\left(T\right)$ is found to be more accurate than those based on $d n_0/d\beta$ 
and the local compressibility. This is understandable because the former approach uses precise 
information of the nature and universality class of the transition in 3D. Nevertheless, as argued 
before, we anticipate that if one decreases the strength of the confining potential and increases 
the number of bosons, as to reach the system sizes that are studied experimentally, then $d n_0/d\beta$ will 
provide accurate results (at least similar to the ones obtained in 2D). This effect is studied  in 
Fig.~\ref{fig:comp3d} where we show the evolution of the critical temperature at the tip of the lobe 
as a function of system size. As the strength of the confining potential is decreased and the  size 
of the system is increased, the estimate of the critical temperature based on $dn_0/d\beta$ tends to 
increase and approach $T_c$ in the homogeneous system. The method based on the local compressibility 
is found to be inadequate close to the tip of the lobe. This is because the maximum of 
$\kappa_{\textrm{diff}}\left(r\right)$ becomes very broad and finite-size effects are stronger. In 
that regime, one also needs a higher accuracy in the determination of the density in order to 
accurately compute the local compressibility. In spite of this, in 3D, the method based on the 
local compressibility yields more accurate results than in 2D (compare Figs.~\ref{fig:comparisons} 
and \ref{fig:comp3d}).  

\section{Conclusions} \label{conclusions}

We have presented a detailed study of the finite temperature phase diagram of strongly correlated bosons 
in the hard-core limit (or the $XY$ model) in two and three dimensions. The critical parameters in the 
homogeneous case were determined through a finite-size scaling analysis of the superfluid stiffness and 
the condensate fraction. We introduced an  approach to estimate the critical temperature from measurements 
of $n_0$ in finite systems. It makes use of the behavior of the derivative $dn_0/dT$ and we derived finite-size 
scaling relations that can be used to extrapolate the results to the thermodynamic limit. This approach can 
be applied to systems that exhibit a diverging zero-momentum occupation in any dimension, irrespective of 
the universality class to which the transition belongs. We showed that this method is also accurate 
in 2D, where the system does not exhibit BEC. Furthermore, we computed the phase diagram using 
mean-field theory and found it to be quantitatively quite different from the results of numerically exact 
QMC simulations in 2D and 3D. Hence, for this model, thermal and quantum fluctuations are strong even 
in three dimensions, and mean-field theory is a poor approximation. 

In the presence of an additional confining potential, we proved that the Bose-Hubbard model does not exhibit 
finite-temperature BEC in two dimensions, provided that density remains finite across the entire system 
in the thermodynamic limit. Moreover, we considered measurements of the critical temperature and 
chemical potential of the homogeneous system based on experimentally measurable quantities such as the 
momentum distribution function and the local density profile. The accuracy of each method discussed 
depends on the dimensionality of the system and the range of temperatures and chemical potentials 
considered. In two dimensions, we found that the approach introduced in this work, based on the 
derivatives of $n_0$ with respect to $\beta$, is accurate in all regions of the phase diagram. 
A method based on the measurement of the local density was found to be reliable when $T_c$ is low, 
while close to the tip of the superfluid lobe this approach is less effective even when the trap is 
very shallow. This can be understood to be due to the strong deviations from the LDA close to the tip of the 
superfluid lobe. A quantitative account of these deviations based on trapped finite-size scaling, as 
presented in Ref.~\onlinecite{ceccarelli_vicari_12} and \onlinecite{ceccarelli_torrero_12}, would in 
principle allow one to perform an accurate size-scaling analysis in the presence of the confining potential, 
which might potentially improve the capabilities of the methods based on the measurements of the density 
profile. The accuracy of the latter method improves in 3D, but still remains inadequate as one approaches 
the tip of the superfluid lobe. In three dimensions, the approach based on $Q\left(T\right)$ was found 
to be the most accurate.

\begin{acknowledgments}
This work was supported by the U.S. Office of Naval Research under Award No.\ N000140910966. 
We thank Nikolai Prokof'ev, Itay Hen, and Rong Yu for useful discussions. 
\end{acknowledgments}

\appendix
\section{\label{sec:minshift} Differences between $dn_0/dT$ and $dn_0/d\beta$}

We briefly illustrate, by means of a simple analysis, why the estimate of the critical temperature 
based on $dn_0/dT$ differs from the estimate based on $dn_0/d\beta$. We also discuss under which 
conditions the two estimates should approach each other. 

We consider $n_0(\beta)$ to be the zero momentum occupation in the vicinity of $\beta_c$. 
Its first derivative, which exhibits a maximum at $\beta^{*}$, can be written as
\begin{equation}
\dfrac{d n_0}{d \beta}= d_0+ a\left( \beta-\beta^{*}\right)^2,
\end{equation}
where the curvature of the parabola is $a<0$, the height of the maximum is $d_0$, and $\beta$ is assumed
to be very close to $\beta^{*}$. If instead we now compute $dn_0/dT$, we anticipate a minimum of this 
function located at a temperature $T^{*}\neq1/\beta^{*}$ given by
\begin{equation}
\dfrac{1}{ T^{*}}= \dfrac{3a\beta^{*} - |a|\sqrt{ \beta^{* 2} -\dfrac{8d_0}{a}}}{ 4 a }.
\end{equation}
In general, the position of the minimum as a function of $T$ depends on the position of the maximum 
$\beta^{*}$, its curvature $a$, and its height $d_0$. However, in the limit of very large system sizes 
and very shallow traps, one expects the maximum of the derivative $dn_0/d\beta$ to be very sharp. 
In our simple example, this regime corresponds to a large value of the curvature, i.e., 
$|d_0/a|\ll \beta^{* 2 }$, which implies that $T^{*} \simeq 1/\beta^{*}$.


\begin{thebibliography}{44}
\bibitem{note:better} Note that, in Ref.~\onlinecite{harada_kawashima_97},
a more sophisticated fitting approach allowed them to obtain the critical value with
a higher accuracy that the one we achieve in here.
\end{thebibliography}
\end{document}